# Particle aggregation by positive dielectrophoresis

Charles Osarinmwian

School of Chemical Engineering and Analytical Science, University of Manchester, Oxford Road, Manchester M13 9PL

Particle aggregates are formed with viable yeast cells (*Saccharomyces cerevisiae*) and bacterial cells (*Escherichia coli* and *Micrococcus luteus*), and are assembled in an interdigitated-castelled microelectrode using positive dielectrophoresis with an alternating current sinusoidal voltage of 2-20 V peak-peak at 1 MHz. The height of *S. cerevisiae* aggregates were increased by the addition of the ampholytes N-[2-hydroxyethyl] piperazine-n'-[2-ethanesulfonic acid] (HEPES) or ε-aminocaproic acid (EACA) due to the resulting rise in permittivity of the aqueous solution. The mean height of *S. cerevisiae* aggregates in the presence of 1 M HEPES was 25.7 % higher than achieved in an ampholyte-free solution. Based on the results obtained, a next-generation triangular grooved microelectrode is proposed as an alternative to the interdigitated-castelled microelectrode. This innovation could mark a paradigm shift from aggregating particles at specific regions of high electric field strength (e.g. electrode edge) towards a layer by layer aggregation of particles along a microelectrode.

The progress of dielectrophoresis as a novel and rapidly growing field has lead to many significant advances in a wide range of diverse applications. Dielectrophorsis arises from induced dipoles in neutral particles exposed to a non-uniform electrical field. In the case of biological cells, the dielectrophoretic force experienced by a cell depends on intrinsic electrical properties such as membrane capacitance and conductance, both of which change with cell type. In tissue engineering, dielectrophoresis can be used to characterize artificial tissues and form artificial tissue-like materials by either assisting in the formation of the artificial extracellular matrix or by the micromanipulation of cells[1]. This is important since hydrated extracellular polymeric substances in nature protect organisms in a biofilm in adverse environments while permitting intense interactions, including cell-cell communication among the cells within a biofilm[2]. The use of dielectrophoresis in bioanalytic microsystems may result in multifunctional platforms for basic biological insights into cells and tissues as well as cell-based sensors with biochemical, biomedical and environmental functions[3]. In this work, the influence of the applied voltage and the presence of ampholytes on the aggregation height of *S. cerevisiae*, *Escherichia coli* and *Micrococcus luteus* in an aqueous-based solution within an interdigitated-castelled microelectrode are investigated.

To obtain particle aggregates, an alternating current sinusoidal voltage of amplitude in the range 2-20 V peak-peak was applied. A single cell particle suspension (consisting of 6-8 µm *S. cerevisiae*, *M. luteus* or *E. coli* cells) was injected on top of the interdigitated-castelled microelectrode assembly using a syringe and then covered with a microscope slide coverslip. *S. cerevisiae* was grown overnight, harvested and washed 4 times in deionized water by repeated centrifugation whereas *M. luteus* (Sigma Aldrich) and *E. coli* (Sigma Aldrich) were used without further treatment. When the particles in the suspension settled in the microelectrode assembly an alternating current sinusoidal voltage was applied at a frequency of 1 MHz. This frequency induces a Maxwell-Wagner-type polarization leading to a strong positive dielectrophoretic force because the cell membrane surrounding the cytoplasm has a low conductivity and permittivity[5]. To investigate the effect of changing solution properties on aggregation height, 50 µL of HEPES (Sigma Aldrich, H4034) or EACA (Sigma Aldrich, A2504) in solution was pipetted into the particle suspension, gently stirred and then covered with a microscope slide coverslip. Aggregate heights were measured using a standard technique that can be found elsewhere[4].

Figure 1 shows that increasing the applied voltage increases the electric field strength around microelectrodes and therefore increases the height of particle aggregates. *M. luteus* has lower cell wall conductivity than *E. coli* leading to a slightly complex particle permittivity that lowers the positive dielectrophoretic force[5] and hence lowers the aggregate height (Fig. 1a). *S. cerevisiae* has the lowest cell wall conductivity and so the lowest aggregate height. The continuity of the displacement flux density across the interface between these particles and solution controls the initial temporal response to the applied electric field due to perturbations of bound charges at the atomic scale[6]. This particle size independent phenomenon may dominate over small variations in particle size despite the positive dielectrophoretic force dependency on the cell size $d$ by a factor $d^3$. Increasing the permittivity of the solution by the addition of ampholytes (i.e. HEPES or EACA) increased the aggregate height (Fig. 1b); the mean height of *S. cerevisiae* aggregate in the presence of 1 M HEPES was 25.7 % higher than achieved in an ampholyte-free solution. The negative effect of the increase in conductivity on the aggregate height is weaker than the positive effect of the increase in permittivity.

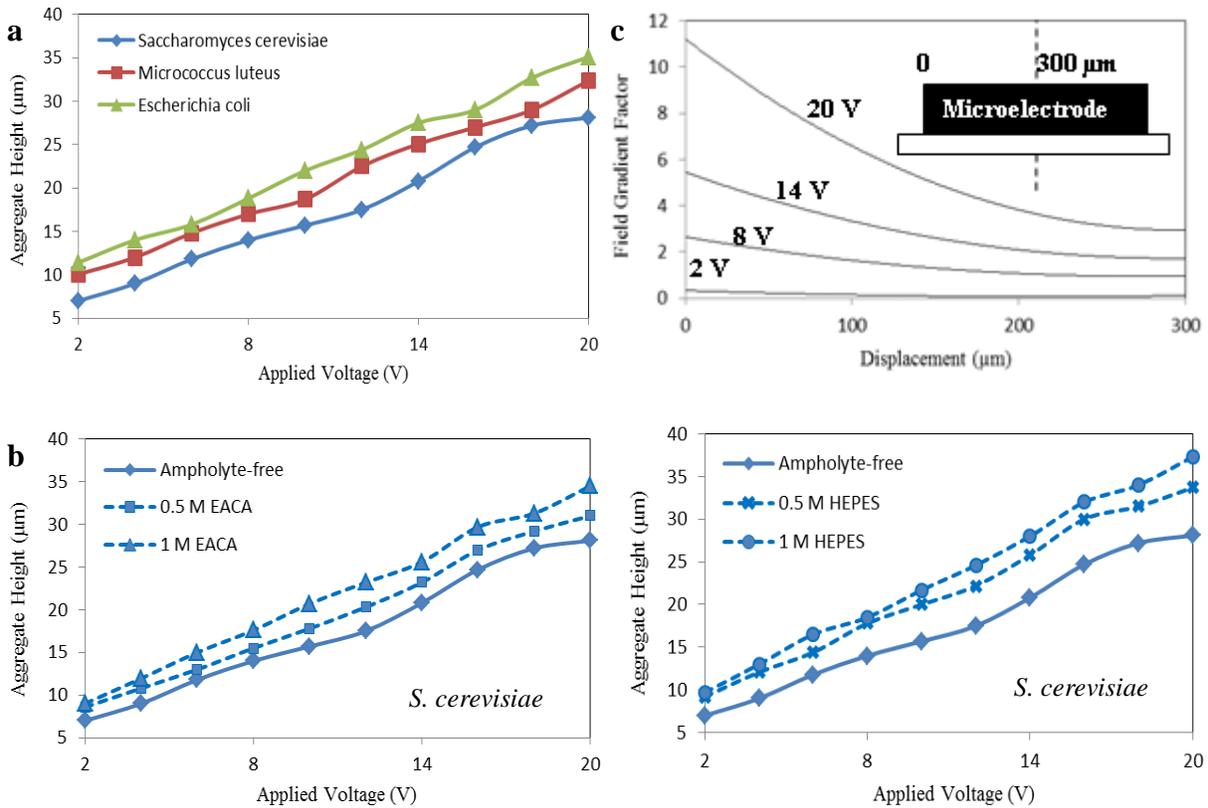

**Figure 1: Aggregation of particles by positive dielectrophoresis.** (**a**) Effect of applied voltage on the aggregate height of different cells. (**b**) Effect of ampholyte addition on the aggregation height of *S. cerevisiae*. (**c**) COMSOL multiphysics simulation results of the magnitude of the field gradient factor along a microelectrode mounted on a substrate. The strongest positive dielectrophoretic force is experienced at the edge of the microelectrode at high applied voltages in solution (electrical conductivity of solution: $2 \times 10^{-4}$ S m$^{-1}$).

The dielectrophoretic force acting on a spherical particle in a non-uniform electric field is dependent on the strength of the effective polarization of the particle (measured by the Clausius-Mossotti factor). In terms of electrostatic energy, a positive value for the real part of the Clausius-Mossotti factor indicates that work is required to withdraw a particle from the highest field region (i.e. positive dielectrophoresis) in which the conduction losses in the particle exhibit an electrical conductivity associated with mobile ions in the particle structure[6]. The current-density vector in the solution is derived from Faraday's law under the assumption of electro-neutrality in the solution (i.e. fairly moderate voltage gradients) and negligible diffusion transport. Given that the electric field strength diminishes radially from the microelectrode edge according to Coulomb's inverse-square law, particles experience strong positive dielectrophoretic forces at the microelectrode edge (i.e. formation of aggregates) (Fig. 1c). It is important to note that increasing the applied voltage is an effective method of increasing aggregate height[5].

The ease of scaling electric fields and developing multifunctional platforms in the next-generation of microelectrode designs may create opportunities for exploiting dielectrophoresis in a range of microsystems. The interdigitated-castelled microelectrode used in this work offers relatively high values of electric field strength at modest applied voltages while allowing for simultaneous observations of both positive and negative dielectrophoresis[4]. However, the shortcomings, in terms of collecting particles at microelectrode edges rather than along a microelectrode, have been addressed using very fine microelectrodes and small working volumes in a zipper-microelectrode[7]. Also, due to geometrical leveling[8], the development of a novel triangular groove microelectrode could collect particles along microelectrode arrays and so induce a layer by layer aggregation of particles. This would require the formation of extracellular polymeric substances between levitated particles (in regions of negative dielectrophoresis) and aggregated particles (in regions of positive dielectrophersis) in order to immobilize the particles into synergistic microconsortia[2]. The current state of affairs indicates that the next-generation of dielectrophoresis technology warrants novel microelectrode designs such as the triangular groove microelectrode.

**Acknowledgements** I thank G. H. Markx for discussions and COMSOL Inc. for providing a license to use their software. This research was funded by EPSRC (UK).